\documentclass[10pt,conference]{IEEEtran}
\IEEEoverridecommandlockouts

\usepackage{cite}
\usepackage{amsmath,amssymb,amsfonts}
\usepackage{algorithmic}
\usepackage{graphicx, wrapfig}
\usepackage{textcomp}
\usepackage{xcolor}
\usepackage{tabularx}
\usepackage{multirow}

\usepackage{picinpar}
\usepackage{hyperref}
\usepackage{paralist}

\newcolumntype{C}{>{\raggedright\arraybackslash}X}
\definecolor{darkred}{rgb}{0.8,0,0}
\def\BibTeX{{\rm B\kern-.05em{\sc i\kern-.025em b}\kern-.08em
    T\kern-.1667em\lower.7ex\hbox{E}\kern-.125emX}}

\newenvironment{bio}[1]
{\par
 \bigskip
 \begin{wrapfigure}{l}[0pt]{1in}
 \vspace{-15pt}
 \includegraphics[height=1in,clip,keepaspectratio]{#1}
 \vspace{-25pt}
 \end{wrapfigure}
 \footnotesize \noindent}
{\par\bigskip}

\begin{document}

\onecolumn

\begin{figure}
    \centering
    \includegraphics[width=.3\textwidth]{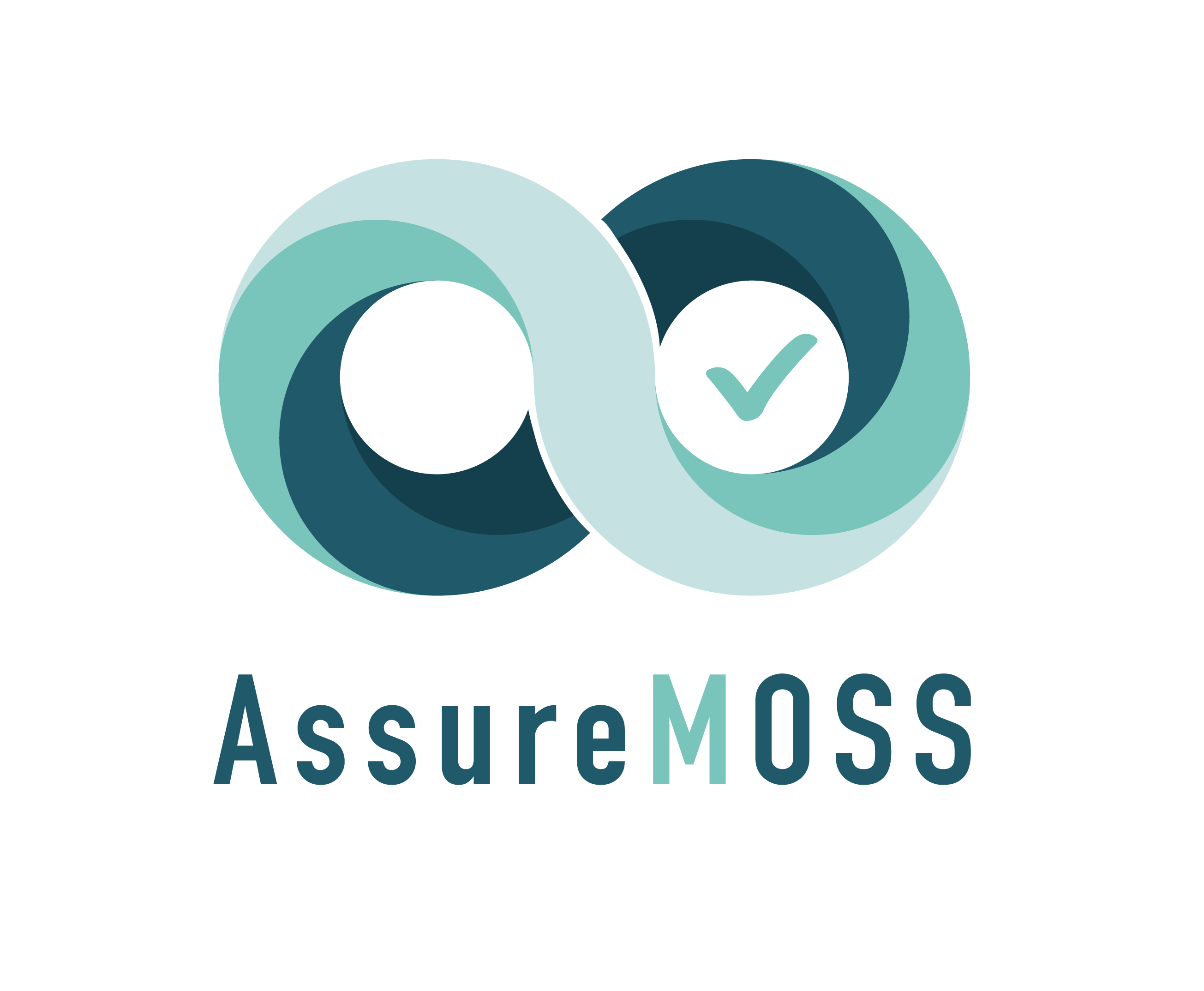}
\end{figure}

\vspace{2\baselineskip}

\begin{figure}[h!]
    \centering
    \includegraphics[width=\textwidth]{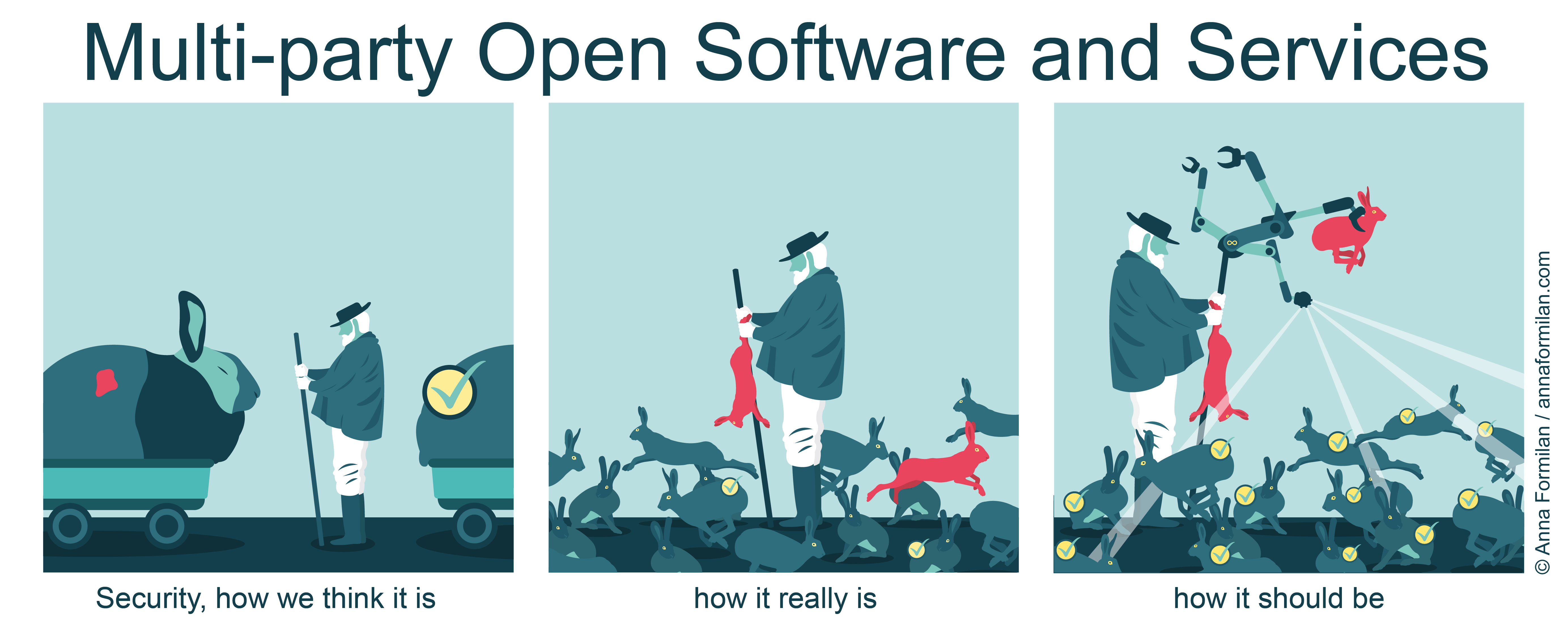}
    % \caption{Caption}
    % \label{fig:my_label}
\end{figure}

\vspace{2\baselineskip}

\begin{center}
{\huge \textbf{Secure Software Development in the Era of Fluid Multi-party Open Software and Services}}
\end{center}

\vspace{\baselineskip}

{\large
Authors:
\begin{itemize}
    \item[] \textbf{Ivan Pashchenko}, University of Trento (IT)
    \item[] \textbf{Riccardo Scandariato}, Hamburg University of Technology (DE)
    \item[] \textbf{Antonino Sabetta}, SAP Security Research (FR)
    \item[] \textbf{Fabio Massacci}, University of Trento (IT), Vrije Universiteit Amsterdam (NL)
\end{itemize}
}

\vfill

\begin{wrapfigure}{l}{2.5cm}
\vspace{-\baselineskip}
\includegraphics[width=2.5cm]{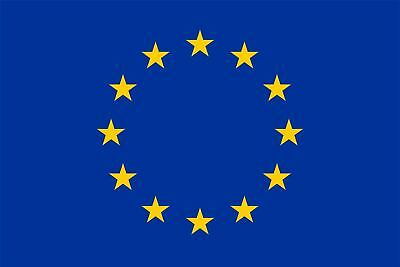}
\end{wrapfigure}

This paper was written within the H2020 AssureMOSS project that received funding from the European Union's Horizon 2020 research and innovation programme under grant agreement No 952647. This paper reflects only the author's view and the Commission is not responsible for any use that may be made of the information contained therein.

\clearpage
\twocolumn
% \vspace*{-6\baselineskip}
\begin{bio}{graph/Logo_colori}
\textbf{Assurance and certification in secure Multi-party Open Software and Services (AssureMOSS)} No single company does master its own national, in-house software. Software is mostly assembled from “the internet” and more than half come from Open Source Software repositories (some in Europe, most elsewhere). Security \& privacy assurance, verification and certification techniques designed for large, slow and controlled updates, must now cope with small, continuous changes in weeks, happening in sub-components and decided by third party developers one did not even know they existed. AssureMOSS proposes to switch from process-based to artefact-based security evaluation by supporting all phases of the continuous software lifecycle (Design, Develop, Deploy, Evaluate and back) and their artefacts (Models, Source code, Container images, Services). The key idea is to support mechanisms for lightweigth and scalable screenings applicable automatically to the entire population of software components by Machine intelligent identification of security issues, Sound analysis and verification of changes, Business insight by risk analysis and security evaluation. This approach supports fast-paced development of better software by a new notion: continuous (re)certification. The project will generate also benchmark datasets with thousands of vulnerabilities. AssureMOSS: \textbf{Open Source Software: Designed Everywhere, Secured in Europe}. More information at \textbf{\url{https://assuremoss.eu}}.
\end{bio}

\begin{bio}{photos/pashchenko}
\textbf{Ivan Pashchenko} (PhD 2019) is a Research Assistant Professor at the University of Trento, Italy. He was awarded a silver medal at the ACM/Microsoft Student Graduate Research Competition at ESEC/FSE. He is UniTrento main contact in \textit{``Continuous analysis and correction of secure code''} work package for the AssureMOSS project. Contact him at \emph{ivan.pashchenko@unitn.it}.
\end{bio}

% \begin{IEEEbiography}[{\vspace{-2\baselineskip}\includegraphics[height=1in,clip,keepaspectratio]{photos/pashchenko}}]{Ivan Pashchenko} (PhD 2019) is a Research Assistant Professor at the University of Trento, Italy. He was awarded a silver medal at the ACM/Microsoft Student Graduate Research Competition at ESEC/FSE. He is UniTrento main contact in \textit{``Continuous analysis and correction of secure code''} work package for the AssureMOSS project. Contact him at \emph{ivan.pashchenko@unitn.it}.
% \end{IEEEbiography}
% \begin{scriptsize}
\vspace*{-\baselineskip}
\begin{bio}{photos/scandariato}
\textbf{Riccardo Scandariato} (PhD 2004) is a professor at the Hamburg University of Technology, Germany. He is the Research Leader of the AssureMOSS project and the leader of a work package on \textit{``Continuous validation of secure design''}.
Contact him at
\emph{riccardo.scandariato@tuhh.de}.
\end{bio}
\vspace*{\baselineskip}
% \begin{IEEEbiography}[{\vspace{-2\baselineskip}\includegraphics[height=1in,clip,keepaspectratio]{photos/scandariato}}]{Riccardo Scandariato} (PhD 2004) is a professor at the Hamburg University of Technology, Germany. He is the Research Leader of the AssureMOSS project and the leader of a work package on \textit{``Continuous validation of secure design''}.
% Contact him at
% \emph{riccardo.scandariato@tuhh.de}.
% \end{IEEEbiography}
% \vspace*{-6\baselineskip}
\begin{bio}{photos/sabetta}
\textbf{Antonino Sabetta} (Phd 2007) is a senior researcher at SAP Security Research. He is the Technical Leader of the AssureMOSS project and the \textit{``Continuous analysis and correction of secure code''} work package leader. Contact him at \emph{antonino.sabetta@sap.com}.
\end{bio}
\vspace*{\baselineskip}
% \begin{IEEEbiography}[{\vspace{-2\baselineskip}\includegraphics[height=0.8in,clip,keepaspectratio]{photos/sabetta}}]{Antonino Sabetta} (Phd 2007) is a senior researcher at SAP Security Research. He is the Technical Leader of the AssureMOSS project and the \textit{``Continuous analysis and correction of secure code''} work package leader. Contact him at \emph{antonino.sabetta@sap.com}.
% \end{IEEEbiography}
% \vspace*{-6\baselineskip}
\begin{bio}{photos/massacci}
\textbf{Fabio Massacci} (Phd 1997) is a professor at the University of 
Trento, Italy, and Vrije Universiteit Amsterdam, The Netherlands. 
He received the Ten Years Most Influential Paper award by the IEEE Requirements Engineering Conference in 2015. 
He is the the European Coordinator of the AssureMOSS project. Contact him at \emph{fabio.massacci@ieee.org}.
\end{bio}
% \begin{IEEEbiography}[{\vspace{-2\baselineskip}\includegraphics[height=1in]{photos/massacci}}]{Fabio Massacci} (Phd 1997) is a professor at the University of 
% Trento, Italy, and Vrije Universiteit Amsterdam, The Netherlands. 
% He received the Ten Years Most Influential Paper award by the IEEE Requirements Engineering Conference in 2015. 
% He is the the European Coordinator of the AssureMOSS project. Contact him at \emph{fabio.massacci@ieee.org}.
% \end{IEEEbiography}
% \end{scriptsize}

How to cite this paper:
\begin{itemize}
    \item Pashchenko, I. and Scandariato, R. and Sabetta, A. and Massacci, F. Secure Software Development in the Era of Fluid Multi-party Open Software and Services. \emph{Proceedings of the International Conference on Software Engineering - New and Emerging Results (ICSE-NIER 2021)}. IEEE Press.
\end{itemize}

License:
\begin{itemize}
\item This article is made available with a perpetual, non-exclusive, non-commercial license to distribute.
\item The graphical abstract is an artwork by Anna Formilan.
\end{itemize}

\clearpage

\title{Secure Software Development in the Era of Fluid Multi-party Open Software and Services}

\author{
\IEEEauthorblockN{Ivan Pashchenko}
\IEEEauthorblockA{\textit{Univ. of Trento}, IT \\
ivan.pashchenko@unitn.it}
\and
\IEEEauthorblockN{Riccardo Scandariato}
\IEEEauthorblockA{\textit{Hamburg Univ. of Technology}, DE \\
riccardo.scandariato@tuhh.de}
\and
\IEEEauthorblockN{Antonino Sabetta}
\IEEEauthorblockA{\textit{SAP Security Research}, FR \\
antonino.sabetta@sap.com}
\and
\IEEEauthorblockN{Fabio Massacci}
\IEEEauthorblockA{\textit{Univ. of Trento}, IT \\ \textit{VU Amsterdam}, NL \\
fabio.massacci@ieee.org}
}
% \author{\IEEEauthorblockN{Author One}
% \IEEEauthorblockA{\textit{Affiliation} \\
% email address}
% \and
% \IEEEauthorblockN{Author Two}
% \IEEEauthorblockA{\textit{Affiliation} \\
% email address}
% \and
% \IEEEauthorblockN{Author Three}
% \IEEEauthorblockA{\textit{Affiliation} \\
% email address}
% \and
% \IEEEauthorblockN{Author Four}
% \IEEEauthorblockA{\textit{Affiliation} \\
% email address}
% }

\maketitle
\thispagestyle{plain}
\pagestyle{plain}

\begin{abstract}
Pushed by market forces, software development has become fast-paced.
As a consequence, modern development projects are assembled from 3rd-party components. Security \& privacy assurance techniques once designed for large, controlled updates over months or years, must now cope with small, continuous changes taking place within a week, and happening in sub-components that are controlled by third-party developers one might not even know they existed. In this paper, we aim to provide an overview of the current software security approaches and evaluate their appropriateness in the face of the changed nature in software development. Software security assurance could benefit by switching from a process-based to an artefact-based approach. Further, security evaluation might need to be more incremental, automated and decentralized. We believe this can be achieved by supporting mechanisms for lightweight and scalable screenings that are applicable to the entire population of software components albeit there might be a price to pay.
\end{abstract}


\section{Security Head-Scratchers}
\label{sec:intro}

In the last decade, three major forces radically changed software development projects. 
\textit{First}, the trends of DevOps as well as continuous integration and delivery have led to a \emph{fast-paced development} with faster product changes~\cite{cooper2016agile,zhang2010agile}.
\textit{Second}, the adoption of agile practices has lead to a \emph{democratised process} where the development is carried out by self-organizing and cross-functional teams that have limited central oversight~\cite{woods2015aligning}. 
%This situation is amplified by the pulverization of software projects into many small parts, e.g., because of the adoption of the micro-service architecture.
%
\textit{Third}, after advocating the benefits of commercial off-the-shelf (COTS) and software reuse for years, we have finally come to the era of \emph{multi-stakeholder development} and software ecosystems, where developers focus on differentiating features in their products and rely on 3rd parties for everything else (e.g, cloud deployment, use of open frameworks, and so on)~\cite{bosch2017}. We observe that the key issue is not that it is Free (albeit this is useful), but that it is in a state of constant changes \emph{and} such changes are done by different people belonging to different organizations.
To capture this phenomenon, we introduce the notion of Multi-party Open Software and Services -- or \emph{MOSS}\footnote{In this paper we are using the term MOSS \emph{differently} from the software metrics domain, where it stands for Measure of Software Similarity~\cite{schleimer2003winnowing}}-- and to reflect the 
dual nature of modern software developer roles, we introduce the notion of \emph{MOSS Prosumer}: to produce MOSS components, software developers not only \textit{develop} their own code, but also \textit{consume} the code from other projects.
%
%The evolution over the past three decades is illustrated in Figure~\ref{fig:soft:past:now}, which refers to the software stack used in the products of SAP. At the end of the 90's, more than 95\% of the software stack in SAP products consisted of their homegrown code. Only databases and operating systems were provided by vendors as licensed closed-source. In contrast, in 2019, the fraction of the homegrown code dramatically decreased to only 5\%: browsers, UI frameworks, package managers, application servers, micro service platforms, containers, containerized operating systems, are all third-party (mostly open-source) software components, with the focus of SAP's own development on differenciating, value-added functionality.
%
%\begin{figure}
%    \centering
%    \includegraphics[width=\columnwidth]{graph/soft-past-now}
%    \caption{Past and new realities in the software stacks (H.~Mack,T.~Schröer, Security Midlife Crisis, SAP Product Security Summit 2019)}
%    \label{fig:soft:past:now}
%\end{figure}
%
From a security perspective, the new situation, brings the two major challenges.

\textbf{Challenge 1 -- Both internal and external sources of insecurity.}
% \noindent\textbf{Challenge 1 -- Securing both internal and external sources.}
Current software security techniques mostly focus on in-house development. The prevalence of the MOSS paradigm means that the root cause of software security problems is shifting from the development of homegrown software to the aggregation of 3rd party (external) components and libraries.
And each actor in the ecosystem might have different security and privacy practices. Security is no longer an internal force.
In CI/CD, on any given sprint (e.g., a two-week development cycle), MOSS prosumers pull in new FOSS libraries, the FOSS community produces security updates for those libraries, MOSS prosumers make decisions that impact the (security) architecture of the systems, new features (and their associated security bugs) are deployed to the customers, and so on. 
This challenges software developers to take additional responsibilities for proper security assessment, security integration, and security maintenance of third-party components. 
I.e., secure coding skills might not be enough.
 
\noindent\textbf{Challenge 2 -- From quality gates to continuous re-certification.} 
As the organisation of software development activities has morphed into a fluid, fast-paced, and democratized process,
the traditional separation in specialized units (e.g., front-end developers, functional testers) that worked under the oversight of software architects and security experts is no longer adequate. 

Rigid ``security gates'' that apply heavyweight security techniques (such as architecture risk analysis, code V\&V) rely on the availability of stable artifacts at fixed points in the development process and assume the supervision of a few highly-skilled security specialists.

% In the ever accelerating software development world, such centralized model is not sustainable anymore, as these ``central experts'' would be overwhelmed by a constantly increasing volume of requests from the development teams.

%Validating, auditing and certifying security is a challenge as traditional methods do not scale.
% Therefore, it is unrealistic to assume that security assurance would be achieved by means of so-called ``security gates'', e.g., by applying heavyweight security techniques (architecture risk analysis, code V\&V) on relatively stable artifacts at fixed points in the development process and through the supervision of a few highly-skilled security specialists.
% In the ever accelerating software development world, such centralized model is not sustainable anymore, as the experts would be overwhelmed by a constantly increasing number of requests from the development teams. 
Such a centralized model needs to evolve to sustain fast-paced software development~\cite{viega2020years}:
\begin{quote}
``If the software security industry is going to make
a meaningful impact, then everything has to be as
easy as humanly possible for both the developers
and users of software.''
\end{quote}

Hence, security assurance might greatly benefit from embedding into the development and operation pipeline in the form of lightweight, intelligent, fully- or semi-automated techniques that can be executed at scale to provide \emph{screening tests} of security-relevant events (e.g., importing an open-source library that requires patching, deploying a container, etc.). These techniques enable an incremental (thus continuous) re-certification of software and their outcome can guide the application of further, in-depth analysis that is more focused and may require expert intervention.
% Hence, there is a need for software security approaches to move towards continuous and incremental re-certification:

\noindent\emph{Scope of this paper} --- The above challenges are cross-cutting and span the entire development process. 
%In this paper, we focus on the development aspects related to the planning and implementation phases as well as the aspect of assessing and certifying software.
With reference to DevSecOps \cite{sec-devops}, we concentrate on the ``Dev'' part; due to space limitations, we leave out most \emph{Ops}-related aspects, which we intend to cover in future work.
\section{Limitations of security assurance techniques}
\label{sec:background}

We discuss the aspects of planning and designing a MOSS project, providing an implementation, and assessing its security with the intent of certification.

\subsection{Secure design}

\emph{Model-driven security.} Several approaches exist to represent security concepts in design models, including architectural models~\cite{van2017design}. Security-aware models are also used in model-driven approaches, e.g., to harden the models via model-to-model transformations and to generate secure implementations via model-to-code transformations~\cite{nguyen2015extensive}.

\emph{Model detection from code.} A general problem with the model-driven security approaches is that, often, there is no guarantee that the model corresponds to the code. Some approaches exist to check the compliance between the planned and implemented security mechanisms~\cite{peldszus2019secure}. These approaches, however, hinge on the presence of an existing security-aware model, which is, in many situations, an unrealistic assumption. As an interesting alternative, security-aware models can be extracted directly from code~\cite{ducasse2009software}.
% The members of the consortium have been active in this field. For instance, in prior work, they aimed to address the problem that most architecture reconstruction techniques suffer from not well reflecting constant evolution of the source in the models~\cite{haitzer2017reconciling}.
However, only few works consider security issues and mostly focus on access control models~\cite{gauthier2011extraction}. Further, fully-automated techniques are notoriously plagued by low levels of precision and recall and produce results that are either not trustworthy or require substantial human checking. In this respect, semi-automated techniques might be more promising~\cite{haitzer2015semi}.

\emph{Model analysis.} State of the practice techniques to assess security at model level are expert-based (e.g., STRIDE~\cite{shostack2014threat} and CORAS~\cite{lund2010model}. Some automated security analysis techniques have been suggested~\cite{berger2016automatically,tuma2019flaws}. However, the design flaws identified by such techniques are confined at the level of the design models: little support is provided to developers with respect to how to modify or refactor their code.

\subsection{Analysis \& correction of secure code}

\emph{Vulnerability detection.} Finding vulnerabilities in software projects is undoubtedly the core software security task. The classic vulnerability finders rely on either static~\cite{chess2004static} or dynamic analysis~\cite{curphey2006web} to identify code locations that are likely to have a weakness. Current analysis tools are known to produce many false positive alerts~\cite{shen2011efindbugs}, which makes them less attractive in practice~\cite{christakis2016developers}. The alternative for code analysis tools are usually referenced as ``defect predictors''~\cite{hall2011systematic}. These techniques employ manually-devised heuristics or machine-learning approaches to construct a model of the source code and then use it to locate a bug or a vulnerability. However, such tools typically work on a file-level of granularity, hence, provide too generic results from practical perspective.
% and this provides limited support to understand which particular code fragments influence their classification decisions.
% to classify a file as potentially buggy or vulnerable.

\emph{Automated corrections of security flaws.} Automatic software repair techniques are proposed to fix bugs. One of the most popular approaches is generate-and-validate based~\cite{gazzola2017automatic}, which first generates a list of candidate patches and verify them using a test suite. The design of repair techniques depends on several factors, such as the selection of modification points or repair operators. These factors strict the success of a technique to specific types of bugs; for example, MutRepair~\cite{debroy2010using} tool is only effective on the bugs related to relational and logical operators while Kali~\cite{qi2015analysis} is preferred for the bugs that can be fixed by only removing code. Thus, designing a new repair approach requires analyzing the characteristics of bugs, which guide the development of the repair components.

\subsection{Secure re-certification of code}

\emph{Dependency studies.} Several studies~\cite{pashchenko2020qualitative,pashchenko2018vulnerable,pashchenko2020vuln4real} argued that current dependency analysis methodologies are based on assumptions that are not valid in an industrial context. They may not distinguish dependency scopes~\cite{kula2017ese} which may lead to reporting non-exploitable vulnerabilities or consider only direct dependencies~\cite{cox2015icse} although security issues may be introduced through transitive dependencies~\cite{lauinger2017thou}.

\emph{Delta Evaluation and Re-certification for Code.} Existing techniques for software security certification~\cite{anisetti2013test} make use of security standards and guidelines provided by certification authorities (e.g., the Common Criteria, NIST, ISO). The industrial need for a lightweight certification scheme is increasing.
% In response to this some solutions
% (from the VESSEDIA\footnote{\url{https://vessedia.eu}} project)
% emerged.
Yet, existing approaches \cite{sec-visa,vessedia} require manual assessments and focus on assessing safety properties and trust in software systems. For instance, Dobbing et al.~\cite{dobbing2005practical} published guidelines for certification of safety properties for COTS Real-Time Operating Systems. Apace with the rapid change in software development practices and changes in the market, source code modifications are often made on legacy, already certified (sub)systems. A fully-fledged re-certification of software security is therefore wasteful. Previous works aim to overcome the problem of high manual effort required to assess security for cloud services. Namely, Anisetti et al.~\cite{anisetti2017semi} presented a certification scheme for the cloud, coupled with a semi-automatic certification process. However, to improve the automation of re-certification, reliable techniques for determining the impacted security from code changes and generating the required documentation for re-certification (i.e., delta evaluation) are needed.
\section{Security Assurance for Fluid MOSS}
\label{sec:solution}

\begin{figure}
    \centering
    \includegraphics[width=\columnwidth]{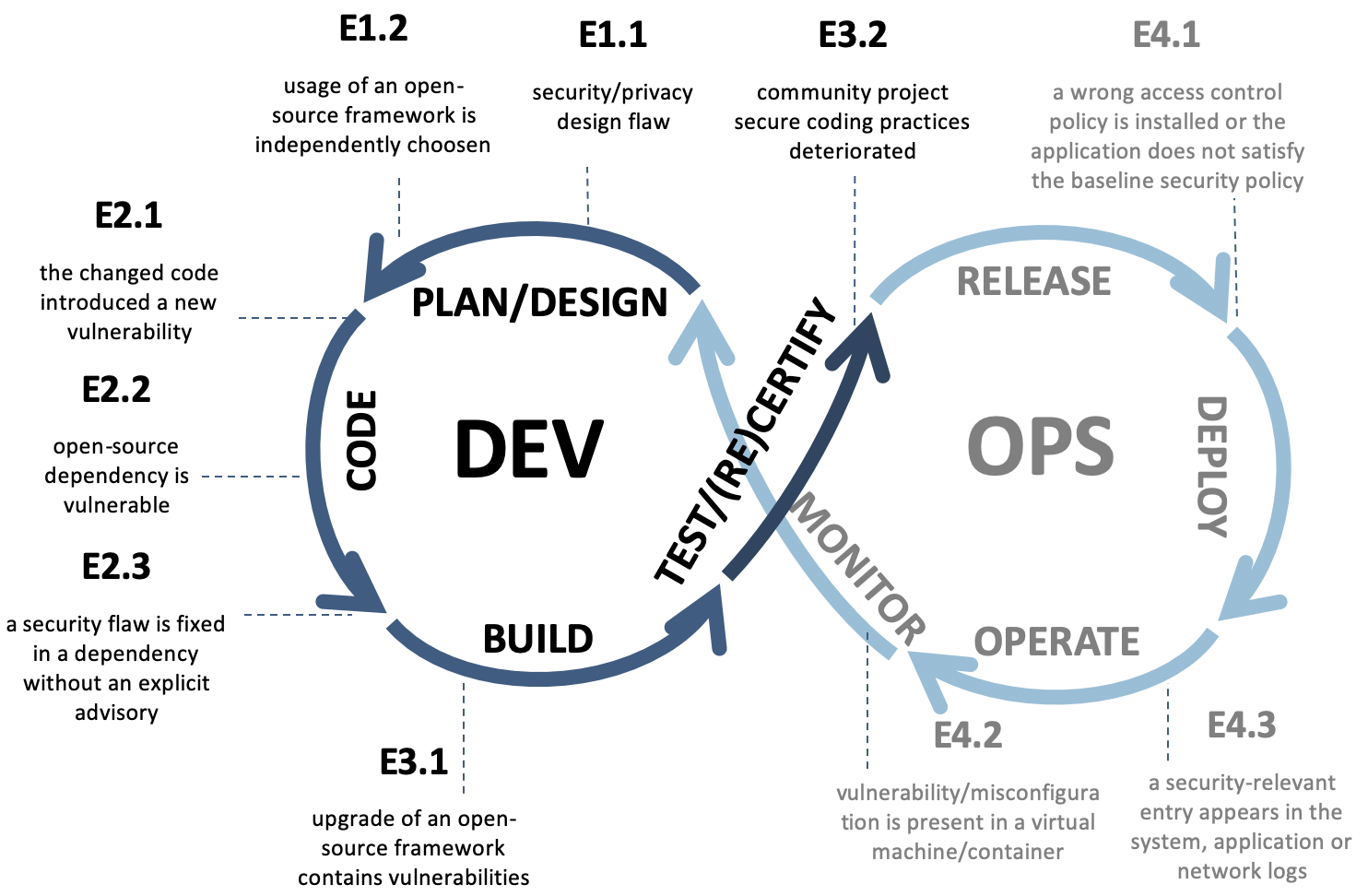}
    \caption{Security events of the MOSS life cycle}
    \label{fig:sec:events}
    \vspace{-10pt}
\end{figure}

\begin{table*}
    \centering
    \caption{Security assurance for Fluid MOSS}
    \label{tab:sec_events}
    \footnotesize
    \begin{tabularx}{\textwidth}{lp{85pt}lp{70pt}XX}
    \hline
\# & Security event & Type & Current mitigation & Limitations in fluid MOSS & Research ideas \\ \hline
\multicolumn{6}{c}{Continuously validated design}\\ \hline
E1.1 & security/privacy design flaw introduced & internal & threat analysis of design model to discover flaw & due to continuously modified/challenged design decisions, the analysis results quickly become obsolete & develop techniques to continuously detect design-level security information in the code \\
E1.2 & FOSS framework chosen & external & security requirements and policies to guide selection & due to the impossibility of central oversight, developers' choices violate the security requirements, which might not even be known & connect security assumptions to code so that the implications of code changes are evident to developers and experts \\ \hline
\multicolumn{6}{c}{Continuous integration of secure code}  \\ \hline
E2.1 & Changed code introduced new vulnerability & internal & static  or  dynamic  analysis  tools & time- and computationally-wise expensive, prone to FPs & lightweight checks that find the code changes introducing vulnerabilities \\
E2.2 & FOSS dependency is vulnerable & external & \multirow{2}{\hsize}{\parbox{1\linewidth}{\vspace{0.3cm} dependency analysis tools}} & \multirow{2}{\hsize}{\parbox{1\linewidth}{\vspace{0.2cm} security analysis is often postponed to the pre-release stage (or even skipped)}} & \multirow{2}{\hsize}{machine learning could be used to analyse patches and automatically generate fixes for vulnerable dependencies} \\
E2.3 & Dependency security flaw is fixed w/o advisory & external &  &  &  \\ \hline
\multicolumn{6}{c}{Incremental evaluation and continuous risk assessment}  \\ \hline
E3.1 & Upgrade of FOSS framework is vulnerable & internal & \multirow{2}{\hsize}{\parbox{1\linewidth}{\vspace{0.5cm} manual assessments with check-lists}} & entire application should be re-certified even if a small part was changed & recertification technique should consider previously certified state and certify only the updated part \\
E3.2 & Community project secure coding deteriorated & external &  & IT risk assessment standards do not provide quantitative measures & a set of risk indicators that cover most parts of the CI/CD pipeline is needed \\
\hline
    \end{tabularx}
    \vspace{-20pt}
\end{table*}

Figure~\ref{fig:sec:events} shows a continuous and high-volume stream of security-relevant events that are generated both internally (because of own development) and externally (because of the dependencies to multi-parties). 
%To enhance security of the DevOps process, Microsoft proposed the following 8 practices (aka, DevSecOps\footnote{\url{https://www.microsoft.com/en-us/securityengineering/devsecops}}):
%\begin{compactenum}
%\item Provide Training
%\item Define Requirements
%\item Define Metrics and Compliance Reporting
%\item Use Software Composition Analysis (SCA) and Governance
%\item Perform Threat Modeling
%\item Use Tools and Automation
%\item Keep Credentials Safe
%\item Use Continuous Learning and Monitoring
%\end{compactenum}
%
Below we provide an outline of the advancements over the state-of-the-art that are necessary in order to cope with these security events.
As space is tyrant, we focus on continuous and incremental design and validation, integration and security assessment of software.
The other Ops-related events, such as wrong access control policy, application violations of baseline security policies, misconfigurations and vulnerabilities present in a virtual machine or container are left to the next paper.

\subsection{Continuous validation of design's security} 
%Here the artifacts of interest are the software architecture and its relation to the MOSS-based code, as well as the run-time system.

\textit{Internal security events.} The software architecture for the project can surface in a proactive way (explicit definition by an architect) or in an emergent way (implicit organization of the code). In both cases, the security tactics chosen by an organisation might introduce a security/privacy design flaw [E1.1]~\cite{santos2017understanding}. For instance, highly sensitive data, like user account data, might be stored in a database in clear-text (confidentiality threat), or there might be no way for the end user to delete their account (privacy threat).

\textit{External security events.} MOSS prosumers might choose to use an open-source framework [E1.2]. However, such choice might jeopardize the integrity of the architecture from a security perspective. For instance, the framework might contain a configuration interface with default credentials that vastly extends the attack surface of the system.

\textit{Current approach.} As mentioned, manual approaches, like model-based threat analysis, do not scale to the MOSS scenario. 
Although there exist automated model analysis approaches, few of them consider security properties and none link models and code. 
Hence, the validation of design from a security point of view is generally performed after the system is already deployed (e.g., penetration tests), and therefore, E1.1 and E1.2 could be discovered only at a later stage when their mitigation is costly and requires significant efforts.

\textit{Research directions in the era of fluid MOSS.} Design artifacts are often abandoned to rust once the implementation work has started, because the code diverges over time from the planned design and this gap nullifies the value of the design models, e.g., for certification. Instead, the design models and their security validation could be incorporated into continuous integration process, by linking design to code. For this, techniques to detect design-level security information in the code should be developed [treatment for E1.1]. To enable the continuous synchronization of the model- and the code-level [treatment for E1.2], these techniques should receive feedback from the (security) experts and support the extraction of design models from the code base.
Finally, there should be automated security checks to validate the design models vis-a-vis the intended security properties, e.g., the confidential management of critical information. 
These checks could be formulated as design-level security tests that can be continuously executed whenever the model of the application changes, e.g., when external components are pulled into the application. 
%The results of the security analysis are oriented towards providing feedback to the development teams and towards supporting the continuous certification of the application. 

% We also incorporate the design models into the process of continuous deployment, by comparing the design models of dev-time artifacts with the behavior of the run-time system. This assures an additional level of correctness for the secure design models, and for the results of the security analysis performed on the models themselves.

\subsection{Continuous integration of secure code}

%The development of MOSS encompasses both internally-developed modules (typically, the differentiating, innovation-oriented code) and components coming from external sources (e.g., FOSS) that are used as  building blocks.

%Ensuring the security of such software and services requires that events spanning the lifecycle of both the internally-developed and the 3rd party components be treated adequately, in a continuous, incremental software development process.

\textit{Internal security events.} While implementing a new feature, developers might introduce a new vulnerability in an internally-developed module that needs to be identified and mitigated before the code is deployed [E2.1]. 

\textit{External security events.} MOSS prosumers have pulled in an open source component to be used in the application under development. 
In turn, the open source component depends on additional (open source) components; its developers (or those of any of its dependencies) might release a new version that could contain vulnerabilities [E2.2]. Unfortunately, MOSS prosumers cannot expect to systematically receive information about security flaws in the 3rd party components of their projects [E2.3]: the practices for reporting vulnerabilities vary across projects and ecosystems, and it is well-known that only a minor part is represented in de-facto vulnerability databases such as the NVD \cite{oss-sec-report-19}, which also suffer from poor accuracy~\cite{dong2019usenix,dashevskyi2018screening}. This makes it difficult for applications that depend on those components to understand the impact of the vulnerability and to correctly determine and compare mitigation strategies (e.g., whether upgrading to a more recent, non-vulnerable version of the component is urgent, or if an application-level change could effectively mitigate the vulnerability at hand and make a potentially expensive version update unnecessary~\cite{ponta2020emse}).

\textit{Current approach.}
Many evaluation studies~\cite{pashchenko2017delta,schaik2020openssf} demonstrate that even industry-grade code analysis tools produce numerous false positive alerts when applied to real-world software projects. In many cases usage of such tools require significant human effort to adjusting the tool to work in the specific environment, while \emph{lightweight} techniques that could work out-of-the-box are only starting to appear. For example, to find a vulnerability in the code of an application, developers have to use static or dynamic analysis tools [treatment of E2.1], that are expensive both time- and computationally-wise, as well as known to generate numerous false positive alerts. Similarly, dependency analysis tools [treatment of E2.2, E2.3] are designed to be run together with code analysis tools. Hence, the code security analysis is often postponed to the pre-release stage (or even skipped altogether), which exponentially increases the cost of fixing discovered vulnerabilities~\cite{dawson2010integrating}. 

\textit{Research directions in the era of fluid MOSS.} In a CI/CD pipeline, unit testing of functionality is automatically performed whenever new code is committed. The outcomes of the unit tests determine whether the code can be integrated. A similar approach could be adapted to security assurance. Automated lightweight checks (leveraging machine learning, where appropriate) are key to identify the code changes that introduce security bugs and need to be reworked [treatment for E2.1]. We believe these checks are crucial to be considered as \emph{screening tests} for vulnerabilities, since they are meant to be applied continuously, automatically, and \emph{quickly}. Code changes should be analyzed automatically and characterized through features that can be efficiently extracted. Such features could include traditional code metrics (e.g., code complexity metrics) as well as properties defined on some abstract representation of the source code (e.g., models based on code tokens, on abstract syntax trees, on control-flow graphs, on data-flow graphs, or on other suitable graph-based representations).

Also, machine learning could facilitate extraction of vulnerability \emph{repair rules} from the FOSS \emph{security patches}. These rules can be automatically applied on commits that are found to introduce a vulnerability, particularly for code that is pulled in from external sources [treatment for E2.2]. Automatic repair might break the code functionality, which is hard to fix for external software, as the application developers are not necessarily familiar with that external code base. Therefore, there is a need for a verification process that selects the relevant test cases to check whether the corrections affected the intended behavior of the application or a MOSS component.  

In this process, the machine learning algorithms must have the fundamental property  of \textit{explainability} (or \textit{interpretability}) of the reason(s) why a particular vulnerability prediction has been made, or a vulnerability repair rule has been suggested.  This feature will aid MOSS prosumers in making informed decisions as to whether or not to accept the change suggestion. However, this is not currently the case. At best ML algorithms suggest that a line of code might contribute with x\% to vulnerabilities, at worst they suggest a file may contain a vulnerabilities. Either way that is not very actionable.

\subsection{Incremental evaluation, certification and risk assessment}
%We tackle the problem of maintaining the risk indicators up to date and preserving the certification status of an application with respect to the development/deployment phases mentioned above.

\textit{Internal security events.} MOSS prosumers have adopted an open source framework and now they want to upgrade to a new version. The upgrade might contain vulnerabilities and could jeopardize the integrity of the application [E3.1]. Re-certifying the entire application every time such upgrades happen is unrealistic.

\textit{External security events.} Considering the previous example, the community project developing a FOSS framework may lose traction (e.g., a key developer has left), so secure coding practices in the project deteriorate [E3.2]. The project manager should be able to assess this trend before the application becomes not sound.

\textit{Current approach.} The automation of existing techniques for software evaluation and certification is limited, therefore, in practice, manual assessments with check-lists for entire software systems still prevail. These techniques suffer from a huge lack of reliability and repeatability due to the non-aligned indicators, over-simplified rating methods and lack of a mandate for certification (who, and according to which standards).
Moreover, the necessity to certify an entire application even if only a small part was modified [E3.1] requires huge amount of expensive effort of security specialists. Additionally, currently no quantitative measures exist when we consider IT risk assessment standards. Hence, nowadays security techniques do not allow project managers to receive operational information about security coding practices of the project [E3.2]. 

\textit{Research directions in the era of fluid MOSS.} To speed up the re-certification of a modified application, one may consider only the code commits that have happened in the application since the previous certification and verify that the changes are not breaking the existing security mechanisms [treatment for E3.1]. To capture early the weakening of security practices, there is a need to have a comprehensive set of risk indicators that cover most parts of the CI/CD pipeline [treatment for E3.2]. These indicators should be updated and analysed periodically (e.g., via time series) during a project life cycle.
\section{Conclusion}

In this paper we have analysed the current state of software security approaches (with focus on software development) and evaluated whether they could efficiently cope with the security events of the MOSS era. 
We argue that software security techniques, once adjusted to the MOSS paradigm, will be beneficial in several ways:
\begin{compactitem}
\item Project managers and security experts will be aware of the security implications made by the MOSS prosumers, e.g., when a FOSS component is used.
\item Security experts will have an `at a glance' overview of the security soundness of the project, e.g., by analyzing the (non-stale) architectural design or by looking at risk indicators that truly reflect the current snapshot.
\item MOSS prosumers will get direct feedback about the security implications of their code commits, both at a feature level (introduction of code vulnerabilities) and at a project level (introduction of architectural flaws).
\end{compactitem}

\medskip
\noindent\textsc{Acknowledgements.} We thank the anonymous reviewers and, in particular, Andy Meneely for their insightful comments. The graphical abstract is an artwork by Anna Formilan \url{http://annaformilan.com}. This work was partially supported by EU-funded project AssureMOSS (grant no.~952647).

\bibliographystyle{IEEEtran}
\bibliography{short-names,biblio}

\end{document}